\documentclass[useAMS,usenatbib]{mn2e}
\usepackage{graphicx}
\newcommand{\oversim}[2]{\protect{\mbox{\lower0.5ex\vbox{%
  \baselineskip=0pt\lineskip=0.2ex
  \ialign{$\mathsurround=0pt #1\hfil##\hfil$\crcr#2\crcr\sim\crcr}}}}}
\newcommand{\simgreat}{\mbox{$\,\mathrel{\mathpalette\oversim>}\,$}} 
\newcommand{\simless} {\mbox{$\,\mathrel{\mathpalette\oversim<}\,$}} 
\title[Maximum stellar mass]
{Evidence for a fundamental stellar upper mass limit from clustered
  star formation}
   \author[C. Weidner and P. Kroupa]{C. Weidner \thanks{E-mail:
   weidner@astrophysik.uni-kiel.de} and P. Kroupa \thanks{E-mail:
   pavel@astrophysik.uni-kiel.de},\thanks{Heisenberg Fellow}\\
   Institut f\"ur Theoretische Physik und Astrophysik, 
              Universit\"at Kiel, 24098 Kiel, Germany}            
\begin{document}

\date{Accepted 28.10.2003}

\pagerange{\pageref{firstpage}--\pageref{lastpage}} \pubyear{2003}

\maketitle

\label{firstpage}

\begin{abstract}
The observed masses of the most massive stars do not surpass about
$150\,M_\odot$. This may either be a fundamental upper mass limit
which is defined by the physics of massive stars and/or their
formation, or it may simply reflect the increasing sparsity of such
very massive stars so that observing even higher-mass stars becomes
unlikely in the Galaxy and the Magellanic Clouds. It is shown here
that if the stellar initial mass function (IMF) is a power-law with a
Salpeter exponent ($\alpha = 2.35$) for massive stars then the richest
very young cluster R136 seen in the Large Magellanic Cloud (LMC)
should contain stars with masses larger than $750\,M_\odot$.
If, however, the IMF is formulated by consistently incorporating a
fundamental upper mass limit then the observed upper mass limit is
arrived at readily even if the IMF is invariant. An explicit turn-down
or cutoff of the IMF near $150\,M_\odot$ is not required; our
formulation of the problem contains this implicitly.  We are therefore
led to conclude that a fundamental maximum stellar mass near
$150\,M_\odot$ exists, unless the true IMF has $\alpha > 2.8$.
\end{abstract}

\begin{keywords}
{stars: early-type -- stars: formation -- stars: luminosity function,
  mass function -- galaxies: star clusters -- galaxies: stellar content}
\end{keywords}

\section{Introduction}
The question on the existence of a finite stellar upper mass limit has
a long history of debate in the literature \citep[][references
therein]{Elme00,Mass98a}.  Observational evidence for such a limit is
scarce because stars more massive than $60-80\,M_\odot$ are very rare.
While stellar formation models lead to a mass limit near
$100\,M_\odot$ imposed by feedback on a spherical accretion envelope
\citep{Kahn74, Wolf86, Wolf87}, theoretical work on the formation of
massive stars through disk-accretion with high accretion rates thereby
allowing thermal radiation to escape pole-wards
\citep[e.g.,][]{Naka89,Jiji96} call the existence of such a limit into
question. Some massive stars may also form by coagulation of
intermediate-mass proto-stars in very dense cores of emerging embedded
clusters driven by core-contraction due to very rapid accretion of gas
with low specific angular momentum, thus again avoiding the
theoretical feedback-induced mass limit \citep{BBZ98, SPH00}.

In his review, \citet{Mass98a} points out that inferring the masses of
very massive stars is difficult due to the fact that stars heavier
than $100\,M_\odot$ do not have their maximum luminosity in the
optical bands and are therefore not easily discriminated on the basis
of photometry from stars with somewhat lower masses.  Using combined
photometric and spectroscopic methods, \citet{Mass98b} find stars with
masses ranging up to $m=140\, M_\odot$ (or even $155\,M_\odot$
depending on the stellar models used) in the rich (about $10^5$~stars)
and very young (1-3~Myr) R136 cluster in the Large Magellanic Cloud,
and that the IMF has a Salpeter exponent ($\alpha=2.35$) for
$3\simless m/M_\odot \simless 100$. Given this IMF, \citet{Mass98a}
emphasises that the observed most-massive-star-mass of around
$150\,M_\odot$ is simply a result of the extreme rarity of even more
massive stars, rather than reflecting a fundamental maximum stellar
mass: the observed numbers of very massive stars are consistent with
the numbers expected from sampling from the IMF and the number of
stars in a cluster.

\begin{figure}
\begin{center}
\includegraphics[width=8cm]{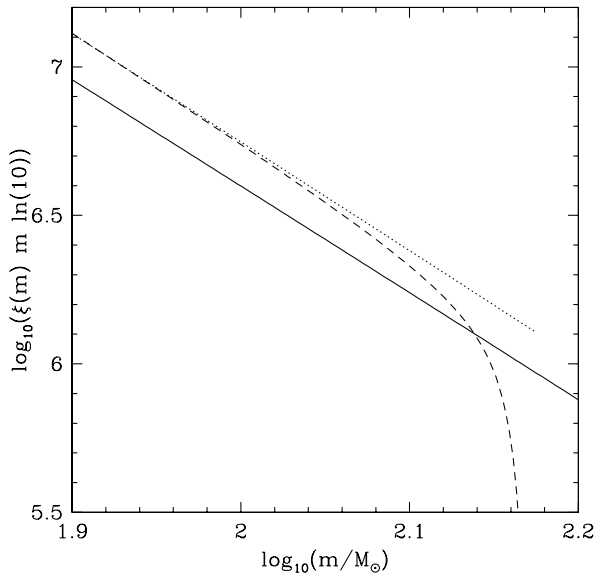}
\vspace*{-2.0cm}
\caption{The ``logarithmic'' IMF ($\xi_{\rm L}(m)=\xi(m) \, m \,
\ln{10}$ over logarithmic stellar mass above $80\,M_\odot$ for three
different cases. The solid line shows an unlimited Salpeter IMF, the
dotted line a Salpeter IMF {\it truncated} at $150\,M_\odot$ and the
dashed line a Salpeter IMF {\it limited} at $150\,M_\odot=m_{\rm
max*}$ in a way described further in \S~\ref{sec:methods}. All three
cases are normalised to the same area over $0.01 \le m/M_\odot <
\infty$.}
\label{fig:imf}
\end{center}
\end{figure}

In order to re-address this last point, we take an approach similar to
the route taken by \citet{Elme00}, but we rely on a different
mathematical formulation.  The idea is to quantify
the expected mass of the most massive star, $m_{\rm max}$, as a
function of the stellar mass, $M_{\rm ecl}$, in an embedded cluster,
and to show that very rich clusters would predict an $m_{\rm max}$
which is significantly larger than the observed most massive star.
Thus we adopt the observed IMF and demonstrate that the observed
cutoff mass is significantly below the expected maximum stellar mass
in rich clusters if there were no fundamental upper mass limit.  The
implication would thus be that there must exist a fundamental upper
mass limit, $m_{\rm max *}$, such that $m_{\rm max} \le m_{\rm max*}$
for all $M_{\rm ecl}$.  With the use of simple equations
concerning the IMF, and the realization that most if not all stars are
born in stellar clusters \citep{Lada03} with an universal IMF, we show
that the solutions of these equations predict a very different high
mass spectrum for a finite or infinite fundamental upper stellar mass,
$m_{\rm max*}$, in dependence of the associated cluster mass. The
principles are shown in Fig.~\ref{fig:imf}.

The next Section~\ref{sec:methods} introduces the equations and the
analytical and numerical methods used to solve them, while the results
are shown in \S~\ref{sec:results}. The implications are discussed in
\S~\ref{sec:discuss}.


\section{Method}
\label{sec:methods}
For our calculations we use a 4-component power-law IMF,

{\small
\begin{equation}
\xi(m) = k \left\{\begin{array}{ll}
\left(\frac{m}{m_{\rm H}} \right)^{-\alpha_{0}}&\hspace{-0.25cm},m_{\rm
  low} \le m < m_{\rm H},\\
\left(\frac{m}{m_{\rm H}} \right)^{-\alpha_{1}}&\hspace{-0.25cm},m_{\rm
  H} \le m < m_{0},\\
\left(\frac{m_{0}}{m_{\rm H}} \right)^{-\alpha_{1}}
  \left(\frac{m}{m_{0}} \right)^{-\alpha_{2}}&\hspace{-0.25cm},m_{0}
  \le m < m_{1},\\ 
\left(\frac{m_{0}}{m_{\rm H}} \right)^{-\alpha_{1}}
    \left(\frac{m_{1}}{m_{0}} \right)^{-\alpha_{2}}
    \left(\frac{m}{m_{1}} \right)^{-\alpha_{3}}&\hspace{-0.25cm},m_{1}
    \le m < m_{\rm max},\\ 
\end{array} \right. 
\label{eq:4pow}
\end{equation}
\noindent with exponents
\begin{equation}
          \begin{array}{l@{\quad\quad,\quad}l}
\alpha_0 = +0.30&0.01 \le m/{M}_\odot < 0.08,\\
\alpha_1 = +1.30&0.08 \le m/{M}_\odot < 0.50,\\
\alpha_2 = +2.30&0.50 \le m/{M}_\odot < 1.00,\\
\alpha_3 = +2.35&1.00 \le m/{M}_\odot,\\
          \end{array}
\label{eq:imf}
\end{equation}}
\noindent where $dN=\xi(m)\,dm$ is the number of stars in the mass
interval $m$ to $m+dm$. The exponents $\alpha_{\rm i}$ represent the
Galactic-field (or standard) IMF \citep{Krou01,Krou02}. The advantage
of such a multi-part power-law description are the easy integrability
and, more importantly, that {\it different parts of the IMF can be
changed readily without affecting other parts}. For example, the
stellar luminosity function for late-type stars poses significant
constraints on the IMF below $m\simless 1\,M_\odot$ \citep{KTG93,
  RGH02, Krou02} which therefore
must remain unaffected when changing the IMF for massive stars.  The
observed IMF is today understood to be an invariant Salpeter power-law above a
few~$M_\odot$, being independent of the cluster density and metalicity
for metalicities $Z \simgreat 0.002$ \citep{Mass98a}.

The basic assumption underlying our approach is the notion that stars
in every cluster follow this same universal IMF.

\subsection{Number of stars}
The Number of stars above a mass $m$ is
\begin{equation}
\label{eq:numb*}
N = \int_{m}^{m_{\rm max*}}\xi(m)~dm,
\end{equation}
when the normalisation constant $k$ (eq.~\ref{eq:4pow}) is given by
the stellar mass of the cluster, 
\begin{equation}
\label{eq:Mecl}
M_{\rm ecl} = \int_{m_{\rm low}}^{m_{\rm max}}m \cdot \xi(m)~dm.
\end{equation}
Here we use the cluster mass in stars prior to gas-blow-out and thus
prior to any losses to the stellar population due to cluster
expansion~\citep{KrouB02}.

\begin{figure}
\begin{center}
\includegraphics[width=8cm]{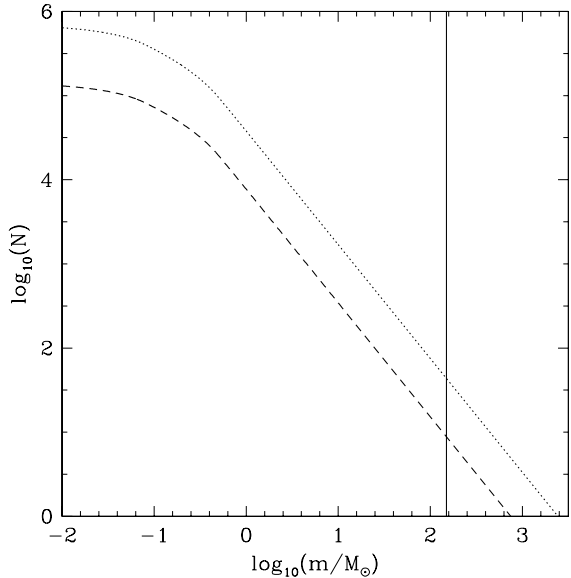}
\vspace*{-2.0cm}
\caption{Number of stars (logarithmic) above mass $m$
  for R136 with different mass estimates \citep[dotted line: $M_{\rm
  R136} = 2.5 \times 10^5\, M_\odot$, dashed line: $M_{\rm R136} = 5
  \times 10^4\, M_\odot$,][]{Sel99}. The vertical solid line marks
  $m=150\,M_\odot$.} 
\label{fig:numbs}
\end{center}
\end{figure}

In Fig.~\ref{fig:numbs} it is shown that a significant number of stars
with masses $m>150\,M_\odot$ should be present in R136 (10 stars
for $M_{\rm R136} = 5 \times 10^4\, M_\odot$ and 40 stars for $M_{\rm
  R136} = 2.5 \times 10^5\, M_\odot$) if no fundamental upper mass
limit exists ($m_{\rm max *} = \infty$) and if the IMF is a Salpeter
power-law above about~$1\,M_\odot$, whereas none are observed. This
sets the problem for which we seek a solution by considering a finite
$m_{\rm max *}$.
\subsection{The limited case}
First we examine the case were a finite upper mass limit for stars
exists. Here two upper mass limits have to be differentiated: the
fundamental maximum-mass a star can have under any circumstances,
$m_{\rm max *}$, and the 'local' upper mass limit $m_{\rm max}\le
m_{\rm max *}$ for stars in a cluster with a stellar mass $M_{\rm
ecl}$. The mass of the heaviest star in a cluster, $m_{\rm max}$,
follows from stating that there is exactly one such star in the
cluster,
\begin{equation}
\label{eq:normlim}
1 = \int_{m_{\rm max}}^{m_{\rm max *}}\xi(m)~dm.
\end{equation}
Note that \citet{Elme00} uses $m_{\rm max *} = \infty$ in his
formulation of the problem.  After inserting eq.~\ref{eq:imf} the
integral can be solved obtaining ($\alpha_{\rm i} \ne 1$):
\begin{equation}
\label{eq:normlim2}
1 = k \cdot \left(\left(\frac{m_{\rm H}}{m_{0}}\right)^{\alpha_{1}}\left(
    \frac{m_{0}}{m_{1}}\right)^{\alpha_{2}}m_{1}^{\alpha{3}}\right) \cdot
    \left( \frac{m_{\rm max *}^{1-\alpha_3}}{1-\alpha_3}-\frac{m_{\rm
    max}^{1-\alpha_3}}{1-\alpha_3}\right),
\end{equation}
as long as $m_{\rm max}>m_1$. For $m_0 \le m_{\rm max} < m_1$ we 
would have
\begin{eqnarray}
\label{eq:normlim3}
1~=&k \cdot \left\{\left(\left(\frac{m_{\rm H}}{m_{0}}\right)^{\alpha_{1}}
m_{0}^{\alpha{2}}\right) \cdot \left(\frac{m_1^{1-\alpha_2}}{1-\alpha_2} - 
\frac{m_{\rm max}^{1-\alpha_2}}{1-\alpha_2}\right)\right.\nonumber\\
+&\left. \left(\left(\frac{m_{\rm H}}{m_{0}}\right)^{\alpha_{1}}\left(
    \frac{m_{0}}{m_{1}}\right)^{\alpha_{2}}m_{1}^{\alpha{3}}\right) \cdot
    \left( \frac{m_{\rm max *}^{1-\alpha_3}}{1-\alpha_3} -
    \frac{m_{1}^{1-\alpha_3}}{1-\alpha_3}\right)\right\}.
\end{eqnarray}
and so on.  For the numerical results obtained in this work $m_{\rm
max*}=150\,M_\odot$ is assumed. 

In order to solve this equation with two unknowns, $k$ and $m_{\rm
max}$, we need another equation. It is provided by the mass in
embedded-cluster stars (eq.~\ref{eq:Mecl}).
With the use of $\xi(m)$ (eq.~\ref{eq:4pow}), eq.~\ref{eq:Mecl} leads
to ($\alpha_{i} \neq 2$):
\begin{eqnarray}
\label{eq:Mecllong}
M_{\rm ecl}~= &k \cdot \left\{ \frac{m_{\rm
    H}^{\alpha_{0}}}{2-\alpha_{0}} \cdot (m_{\rm H}^{2-\alpha_{0}}-m_{\rm
    low}^{2-\alpha_{0}})\right.\nonumber\\
+&\frac{m_{\rm H}^{\alpha_{1}}}{2-\alpha_{1}} \cdot
    (m_{0}^{2-\alpha_{1}}-m_{\rm H}^{2-\alpha_{1}})\nonumber\\
+&\frac{\left(\frac{m_{\rm
    H}}{m_{0}}\right)^{\alpha_{1}}m_{0}^{\alpha{2}}}{2-\alpha_{2}}
    \cdot (m_{1}^{2-\alpha_{2}}-m_{0}^{2-\alpha_{2}})\nonumber\\
+&\left.\frac{\left(\frac{m_{\rm H}}{m_{0}}\right)^{\alpha_{1}}\left(
    \frac{m_{0}}{m_{1}}\right)^{\alpha_{2}}m_{1}^{\alpha{3}}}{2-\alpha_{3}} 
    \cdot (m_{\rm max}^{2-\alpha_{3}}-m_{1}^{2-\alpha_{3}})\right\}
\end{eqnarray}
for $m_{\rm max} > m_{1}$ and with $m_{\rm low}$ set to $0.01\,
M_\odot$ in what follows. For $m_{0} \le m_{\rm max} < m_{1}$ 
eq.~\ref{eq:Mecllong} would be truncated at an earlier term, and so on.

Finally inserting eq.~\ref{eq:normlim2} after a short transformation
into \ref{eq:Mecllong} gives $M_{\rm ecl}$ in dependence of $m_{\rm
max}$. This must now be solved for $m_{\rm max}$ in dependence of
$M_{\rm ecl}$. This is done by finding the roots of this result after
subtracting $M_{\rm ecl}$. Fig.~\ref{fig:limunlim} shows the solution
for a power-law with $\alpha_{3} = 2.35$ and $m_{\rm max *}=150\,
M_\odot$ as a dashed line.

\subsection{The unlimited case}

In the case of $m_{\rm max *}=\infty$ eqs~\ref{eq:Mecl} and
\ref{eq:Mecllong} remain as they are while \ref{eq:normlim} and
\ref{eq:normlim2} change to
\begin{equation}
\label{eq:normunlim}
1 = \int_{m_{\rm max}}^{\infty}\xi(m)~dm
\end{equation}
and (as long as $m_{\rm max}>m_1$ and $\alpha_{3} > 1$)
\begin{equation}
\label{eq:normunlim2}
1 = - k \cdot \left(\left(\frac{m_{\rm H}}{m_{0}}\right)^{\alpha_{1}}\left(
    \frac{m_{0}}{m_{1}}\right)^{\alpha_{2}}m_{1}^{\alpha{3}}\right) \cdot
    \left( \frac{m_{\rm max}^{1-\alpha_3}}{1-\alpha_3}\right),
\end{equation}
respectively.  As only the normalisation factor $k$ deduced from
\ref{eq:normunlim2} changes, eq.~\ref{eq:Mecllong} stays the same, and
inserting \ref{eq:normunlim2} into \ref{eq:Mecllong} gives $M_{\rm
ecl}$ in dependence of $m_{\rm max}$ for the unlimited case.

\begin{figure}
\begin{center}
\includegraphics[width=8cm]{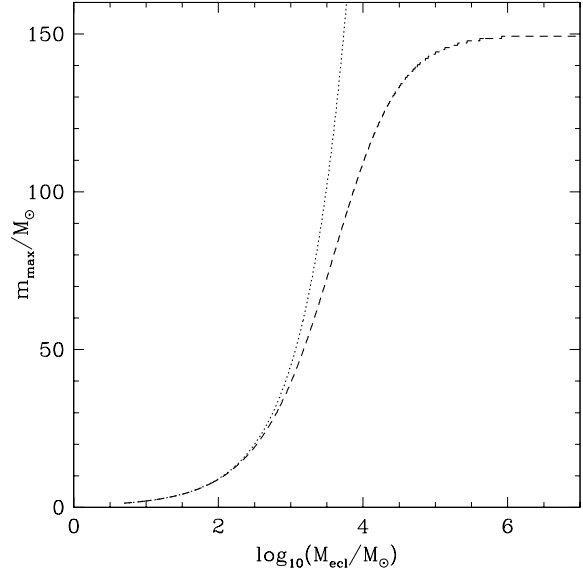}
\vspace*{-2.0cm}
\caption{Dependence of the stellar upper mass limit, $m_{\rm max}$, on
  the cluster mass for a limited ($m_{\rm max *}=150\,M_\odot$;
  dashed line) and an unlimited ($m_{\rm max *}=\infty$;
  dotted line) fundamental upper stellar mass
  and $\alpha_3 = 2.35$.}
\label{fig:limunlim}
\end{center}
\end{figure}

Fig.~\ref{fig:limunlim} shows that the solution for unlimited stellar
masses (dotted line) has a much faster rise than the limited case. If
there were no fundamental upper mass limit for stars then a Salpeter
IMF would predict stars with much larger masses ($m_{\rm
max}>200\,M_\odot$) for clusters with $M_{\rm ecl}>10^{4.5}\, M_\odot$
than are observed to be present. This is also found to be the case by
\citet{Elme00}.


\section{Results}
\label{sec:results}

\begin{figure}
\begin{center}
\includegraphics[width=8cm]{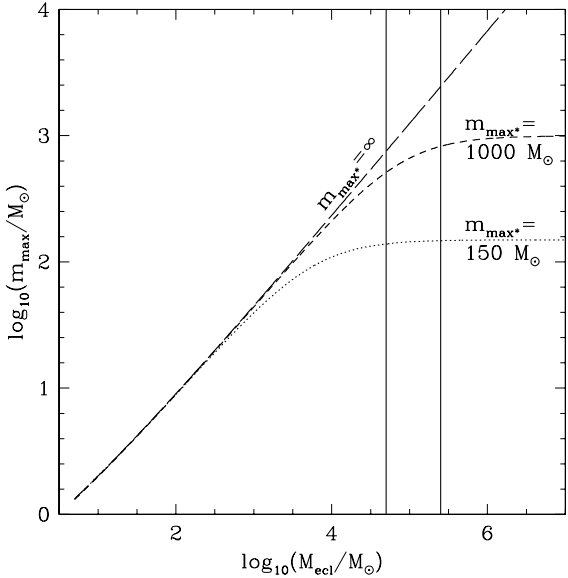}
\vspace*{-2.0cm}
\caption{Double logarithmic plot of the maximal stellar mass versus
  cluster mass. Shown are three cases: finite total upper mass limit
  of $m_{\rm max*}=150\, \, M_\odot$ (dotted line), $m_{\rm
  max *}=1000\, M_\odot$ (short-dashed) and no limit, $m_{\rm max *}=\infty$
  (long-dashed). The vertical lines mark the empirical mass interval for
  R136 in the LMC.}
\label{fig:mmaxmecl}
\end{center}
\end{figure}

\begin{figure}
\begin{center}
\includegraphics[width=8cm]{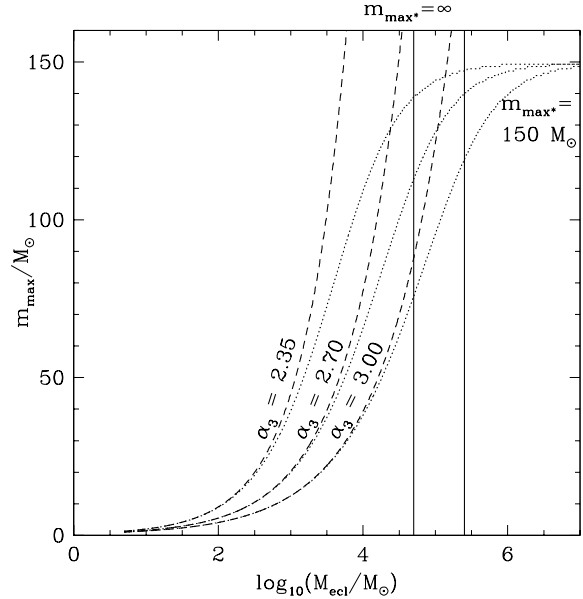}
\vspace*{-2.0cm}
\caption{Maximal stellar mass versus cluster mass
(logarithmic). Results are shown for different exponents ($\alpha_3$)
above $1\,M_\odot$ and for the limited ($m_{\rm max*}=150\,
M_\odot$) and unlimited case. The vertical lines mark the empirical
mass interval for R136 in the LMC.}
\label{fig:mmameexp}
\end{center}
\end{figure}

The results of solving $m_{\rm max}(M_{\rm ecl})$ for a grid of
cluster masses ranging from $M_{\rm ecl}=5\,M_\odot$
(Taurus--Auriga-like stellar groups) to $10^{7}\,M_\odot$ (very
massive stellar super clusters) are plotted in Figs~\ref{fig:mmaxmecl}
and \ref{fig:mmameexp}.

Fig.~\ref{fig:mmaxmecl} shows the variation of the maximum possible
mass for a star, $m_{\rm max}$, in dependence of the cluster mass,
$M_{\rm ecl}$. In the unlimited case (long-dashed line) a linear
relation (in double logarithmic units) is seen. Two vertical lines
indicate the observational mass interval for R136 in the Large
Magellanic Cloud \citep{Sel99}. Without a fundamental upper mass limit
R136, for which~\citet{Mass98b} measure a Salpeter power-law IMF for
$m>\,$few$\,M_\odot$, should have stars with $m>750\, M_\odot$,
whereas no stars with $m>150\,M_\odot$ are seen. Similar values are
found from statistical sampling of the IMF \citep{Elme00}.  For
$m_{\rm max*}=150\,M_\odot$ (dotted line), on the other hand, the
cluster has an upper limit of $140-150\,M_\odot$, in agreement
with the observational limit.

The influence of the high-mass exponent $\alpha_{3}$ on the $m_{\rm
max}(M_{\rm ecl})$ relation is shown in Fig.~\ref{fig:mmameexp}.
Plotted are graphs for limited ($150\,M_\odot$) and unlimited cases,
each for $\alpha_{3} = 2.35$ (Salpeter), 2.70 and 3.00. {\it Exponents
$\alpha_3>2.8$ lead to a $m_{\rm max}(M_{\rm ecl})$ relation which
allows upper masses in R136 of around $150\,M_\odot$ even for the
unlimited case $(m_{\rm max *}=\infty)$}.  Fig.~\ref{fig:almm} shows
that in the case of R136 and for $\alpha_3>2.8$ no distinction can be
made between $m_{\rm max *}=150\,M_\odot$ and $\infty$ given the
uncertainty in $M_{\rm ecl}$.

\begin{figure}
\begin{center}
\includegraphics[width=8cm]{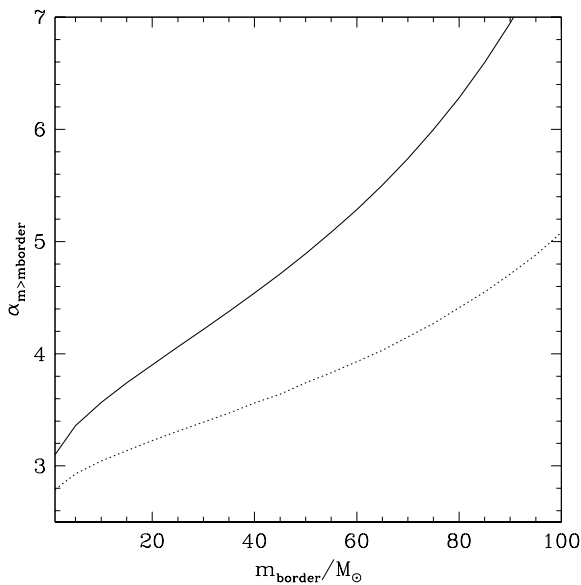}
\vspace*{-2.0cm}
\caption{The power-law exponent $\alpha$ needed to produce a high-mass
  limit of $150\,M_\odot$ for R 136 (solid line: $M_{\rm R136} = 2.5
  \times 10^5\, M_\odot$ and dotted line: $M_{\rm R136} = 5 \times 10^4\,
  M_\odot$) when the IMF is Salpeter up to a certain mass limit
  $m_{\rm border}$.}
\label{fig:turndown}
\end{center}
\end{figure}

Because massive stars are very rare the IMF exponent is often based on
limited statistics and usually only for stars with $m\simless
40\,M_\odot$.  We therefore also consider now the possibility that the
IMF slope is Salpeter to a certain limit (e.g. $40\,M_\odot$) but then
turns down sharply. For this purpose we set $m_1 = m_{\rm border}$ in
eq.~\ref{eq:4pow} with $\alpha_2 = 2.35$ (0.5 $M_\odot$ -- $m_{\rm
  border}$) and find that $\alpha_{m > m_{\rm border}} = \alpha_3$
such that eq.~\ref{eq:normunlim} is fulfilled for $m_{\rm max} =
150\,M_\odot$. The result is plotted in Fig.~\ref{fig:turndown}.

From Fig.~\ref{fig:turndown} it is evident that in order to reproduce
the observed limit of about 
$150\,M_\odot$ for R136 from a formally unlimited mass-scale and a
down-turn mass ($m_{\rm border}$) of, say, $40\,M_\odot$ the exponent
has to change to $\alpha_{m > m_{\rm border}}=3.6$ (for $M_{\rm
  R136} = 5 \times 10^4\, 
M_\odot$) or~4.5 ($M_{\rm R136} = 2.5 \times 10^5\, M_\odot$). Such a
down-turn near $40\,M_\odot$ is not seen in those populations that do
contain more-massive stars \citep[e.g. R136 contains about 40~O3
  stars,][]{Mass98b}, and we therefore consider $m_{\rm max*}\approx
150\,M_\odot$ as being the more realistic possibility. Note though
that the existence of $m_{\rm max*}$ leads to a sharp decline of the
IMF near $120\,M_\odot$ which leads to a similar effect as an increase
of $\alpha_{m > m_{\rm border}}$ near this mass
(Fig.~\ref{fig:imf}). However, our 
formulation needs one additional parameter ($m_{\rm max*}$) to
implicitly account for this turn-down of the IMF, while modelling an
explicit turn-down would need two additional parameters ($m_{\rm
  border}$ and $\alpha_{m > m_{\rm border}}$). 

For massive stars the multiplicity proportion is typically very high
with most O~stars having more than one companion
\citep[e.g.][]{Zinn03, Krou03}
possibly implying the true underlying binary-corrected IMF to have
$\alpha_{3} \simgreat 2.7$ \citep[][Weidner \& Kroupa,
in preparation]{SaRi91}. If this is the case then $m_{\rm max*}$
cannot be 
constrained given the available stellar samples because the Local
Group does not contain sufficiently massive, young clusters.

{\bf
\begin{figure}
\begin{center}
\includegraphics[width=8cm]{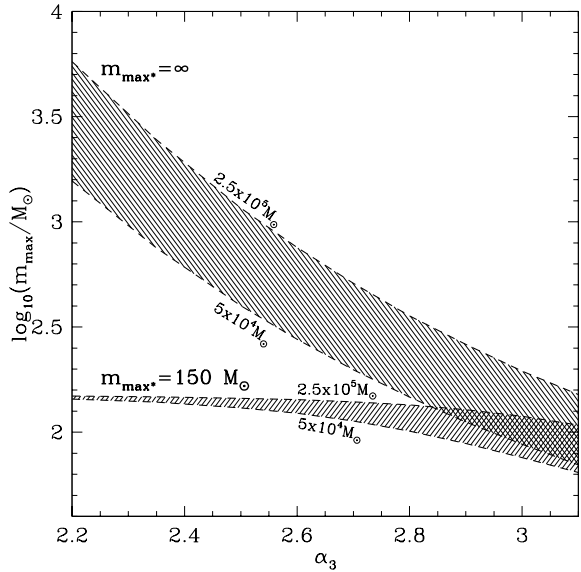}
\vspace*{-2.0cm}
\caption{The mass limits ($m_{\rm max}$) in dependence of the IMF
exponent $\alpha_{3}$ (above $1\, M_\odot$) in the limited case
($m_{\rm max*}=150\,M_\odot$) and the unlimited case ($m_{\rm
max*}=\infty$) for the two mass limits of R136 shown in
Figs.~\ref{fig:mmaxmecl} and \ref{fig:mmameexp}.}
\label{fig:almm}
\end{center}
\end{figure}}

\section{Discussion and Conclusions}
\label{sec:discuss}

With a rather simple formalism based on the current knowledge of the
IMF we have shown that the mere existence of a fundamental upper mass
limit implies the highest mass a star can have in a massive cluster to
be different to the case without such a limit. For low-mass clusters
($M_{\rm ecl} < 10^{3}\, M_\odot$) the differences of the solutions
are negligible (Fig.~\ref{fig:mmaxmecl}), but in the regime of the
so-called 'stellar super-clusters' 
($M_{\rm ecl} > 10^{4}\, M_\odot$) they become very large. Without
such a limit, clusters like R136 in the LMC would have stars with
$m>750\,M_\odot$. 

\citet{Elme00} presents a random sampling model for star formation
from the IMF which is similar to our model. However, Elmegreen
assumes a Salpeter power-law IMF above $0.5\,M_\odot$ and no specific
stellar mass limit. In order to reduce the number of high-mass stars
above $\sim 130\, M_\odot$ he assumes an exponential decline for the
probability to form a star after a turbulent crossing time. The
results of the \citet{Elme00} model are summarised by him as follows:
``There is a problem getting both the Salpeter function out to
$\sim 130\, M_\odot$ in dense clusters and at the same time not
getting any $\sim 300\, M_\odot$ stars at all in a whole galaxy.''

He discusses the following six explanations for this problem:
\begin{itemize}
\item[i.] Stars more massive than $\sim 150\, M_\odot$ exist but have
  not been found yet.
\item[ii.] A self-limitation in the star formation process prohibits stars
  above a certain limit.
\item[iii.] Super-massive stars exist but evolve so quickly that they do not
  leave their primordial clouds -- making them observable only as
  ultra-luminous infrared sources.
\item[iv.] An assumed limit of the cloud size for coherent star formation
\item[v.] The star forming clouds are destroyed after a star of a certain
  (maximum) mass forms.
\item[vi.] The IMF is not universal but different for various star
  forming regions.
\end{itemize}

Case~i can be excluded here because of the number of super-massive
stars expected, for example in R136. Concerning case~iii no such
sources have been found to our knowledge.  The cases ii, iv and v lead
to a physical upper limit consistent with this work. From the point of
view of this work it is not possible to differentiate between
them. Finally as several observations of various clusters show a
universal Salpeter IMF up to $\sim 120\,M_\odot$
\citep[e.g.][]{Mass98b,Sel99,SmGa01} case vi appears
unlikely. Elmegreen thus sees the finite upper mass limit as a cut-off
to the unlimited solution. 

In contrast, we introduce the fundamental upper mass limit
consistently into the formulation of the problem, and together with
the use of a realistic IMF we are able to show strong deviations
of the solutions beyond a simple cut-off. The formulation presented
here has the advantage of explaining the observations under the rather
simple notion that all stars form with the same universal IMF.


\section*{Acknowledgements}
This work has been funded by DFG grants KR1635/3 and KR1635/4.

\bsp

\label{lastpage}
\end{document}